\documentclass[a4paper,11pt]{article}
\usepackage{pos}

\newcommand{\RNum}[1]{\uppercase\expandafter{\romannumeral #1\relax}}
\usepackage{hyperref}
\hypersetup{colorlinks=true,linkcolor=blue,citecolor=blue,urlcolor=blue}
\makeatletter
     {\bgroup\raggedright\small\section*{\refname
        \@mkboth{\MakeUppercase\refname}{\MakeUppercase\refname}}%
      \list{\name{bib\@arabic\c@enumiv}
            \@biblabel{\@arabic\c@enumiv}}%
           {\settowidth\labelwidth{\@biblabel{#1}}%
            \setlength\itemsep{2pt}
            \setlength\parskip{0pt}
            \setlength\parsep{0pt}
            \setlength\partopsep{0pt}
            \leftmargin\labelwidth
            \advance\leftmargin\labelsep
            \@openbib@code
            \usecounter{enumiv}%
            \let\p@enumiv\@empty
            \renewcommand\theenumiv{\@arabic\c@enumiv}}%
      \sloppy\clubpenalty4000\widowpenalty4000%
      \sfcode`\.\@m}
     {\def\@noitemerr
       {\@latex@warning{Empty `thebibliography' environment}}%
      \endlist\egroup}
\makeatother

\title{Extracting \texorpdfstring{$B_s\to D_s^*\ell\nu_\ell$}{Bs -> Ds*l nu}
form factors }

\author*[a]{Anastasia Boushmelev}
\author[b]{Matthew Black}
\author[a]{Oliver Witzel}
\onbehalf{\newline (RBC/UKQCD collaborations)}
\affiliation[a]{Theoretische Physik 1, Center for Particle Physics Siegen, Naturwissenschaftlich-Technische Fakult\"at, Universit\"at Siegen, 57068 Siegen, Germany}

\affiliation[b]{Higgs Centre for Theoretical Physics, School of Physics and Astronomy, University of Edinburgh, Edinburgh EH9 3JZ, UK}

\emailAdd{Anastasia.Boushmelev@uni-siegen.de}

\abstract{Semileptonic $B_{(s)}$ decays are of great phenomenological interest because they allow to determine e.g. CKM matrix elements or test lepton flavor universality.
Taking advantage of already existing lattice data, we demonstrate the analysis steps to extract the four form factors describing exclusive semileptonic
$B_s\to D_s^*\ell\nu_\ell$
decays using the narrow width approximation. Our data are based on RBC/UKQCD's set of 2+1 flavor gauge field ensembles with Shamir domain-wall fermion and Iwasaki gauge field action featuring inverse lattice spacings of $a^{-1}=$1.785, 2.383, and 2.785 GeV as well as pion masses between 268 and 433 MeV. Light, strange and charm quarks are simulated using domain-wall fermions, whereas bottom quarks are generated with the relativistic heavy quark (RHQ) action.}

\FullConference{The 42nd International Symposium on Lattice Field Theory (LATTICE2025)\\
2-8 November 2025\\
Tata Institute of Fundamental Research, Mumbai, India\\}

\begin{document}
\maketitle

\section{Introduction}
Precision tests of the flavor sector in the standard model (SM) are a promising avenue to discover signs of \emph{new physics}\/ not described by the SM. Due to their large mass, processes involving bottom quarks are of particular interest because the large $b$-quark mass allows for many decay channels and the decay may also provide the needed energy to couple to heavy particles not contained in the SM. 
In order to test the SM, we require both precise experimental measurements (e.g.~carried out by the $B$ factories BaBar, Belle or Belle II or by the experiments at the Large Hadron Collider (LHC), LHCb, Atlas and CMS) as well as precise theoretical predictions which are commonly dependent on hadronic matrix elements parametrized by e.g.~decay constants, bag parameters, or form factors.

Focusing on semileptonic decays with $b\to c$ quark transitions, we highlight two long-standing puzzles attracting the attention of the particle physics community:
\begin{enumerate}
\item Extracting the Cabbibio-Kobayashi-Maskawa (CKM) matrix element $|V_{cb}|$ exhibits a tension between \emph{inclusive}\/ determinations summing over all semileptonic final states containing a hadron with a charm quark and  \emph{exclusive} decay channels where the hadronic final state is explicitly specified. Unfortunately, this tension does not have a good explanation in terms of a new physics scenario.  
\item Ratios testing the universality of lepton flavors in weak interactions
\begin{align}  
\mathcal{R}(D^{(*)})=\dfrac{\mathcal{B}(B\to D^{(*)} \tau \nu_\tau)}{\mathcal{B}(B \to D^{(*)} l\nu_l)}
\qquad\text{with}\quad l={e,\mu},
\end{align}
show a tension between experimentally and theoretically determined values. These ratios are theoretically clean because the ambiguity due to $V_{cb}$ as well as other systematic effects cancel. While several new physics scenarios are discussed in the literature, the magnitude of the new physics is surprisingly large for a process occurring at tree-level in the SM. 
\end{enumerate}
Hence further input on the determinations underlying the extraction of $|V_{cb}|$ or the ${\cal R}$-ratio is welcome.
Focusing on semileptonic $B_{(s)}\to D_{(s)}^*\ell\nu$ decays where the hadronic final state is a spin-1 vector particle, we note that form factors over the full $q^2$ range have been published by Fermilab/MILC \cite{FermilabLattice:2021cdg}, JLQCD \cite{Aoki:2023qpa} and HPQCD \cite{Harrison:2021tol,Harrison:2023dzh}, while experimental measurements have been reported by BaBar \cite{BaBar:2007cke, BaBar:2007ddh,BaBar:2025update}, BELLE \cite{Belle:2018ezy,Belle:2019rba}, BELLE \RNum{2} \cite{Belle-II:2020ewe,Belle-II:2020dyp} and LHCb \cite{LHCb:2015gmp,LHCb:2017smo,LHCb:2022piu}.
 In the following we report updates on our efforts to determine the four form factors describing semileptonic $B_{(s)}$ decays to hadronic vector final states where we treat the vector final state as a QCD-stable particle. Previously we reported on this project in Ref.~\cite{Boushmelev:2024jqg} and further details can be found in Ref.~\cite{Boushmelev:Thesis26}.  In the case of $D_{(s)}^*$ mesons created in $b\to c$ transitions this is a good approximation because the $D_{(s)}^*$ width is very small, below 2.1 MeV \cite{ParticleDataGroup:2024cfk}. 
The vector and axial-vector matrix elements arising in the SM decay process can be parametrized in terms of one vector and three axial form factors, reading
\begin{align}
  \langle D_{(s)}^*(k,\varepsilon) |\bar c \gamma^\mu b| B_{(s)}(p)\rangle &= V(q^2) \frac{2\mathfrak{i}\epsilon^{\mu\nu\rho\sigma}\varepsilon_\nu^*k_\rho p_\sigma}{M_{B_{(s)}}+M_{D^*_{(s)}}}, 
  \end{align}
  \begin{align}
        \langle D_{(s)}^*(k,\varepsilon) |\bar c \gamma^\mu\gamma_5 b| B_{(s)}(p)\rangle&= A_0(q^2)\frac{2M_{D^*_{(s)}} \varepsilon^*\cdot q}{q^2}   q^\mu \nonumber\\
    &\quad+ A_1(q^2)\left(M_{B_{(s)}} + M_{D^*_{(s)}}\right)\left[ \varepsilon^{*\mu}  - \frac{\varepsilon^*\cdot q}{q^2}   q^\mu \right] \nonumber\\
    &\quad-A_2(q^2) \frac{\varepsilon^*\cdot q}{M_{B_{(s)}}+M_{D^*_{(s)}}} \left[ k^\mu + p^\mu - \frac{M_{B_{(s)}}^2 -M_{D^*_{(s)}}^2}{q^2}q^\mu\right],\label{eq:ff}
\end{align}
where we use the relativistic convention for the form factors $V$, $A_0$, $A_1$, and $A_2$. 
The momentum and polarization of the $D_{(s)}^*$ is denoted by $k$ and $\varepsilon$, the $B_{(s)}$ momentum is labeled $p$, and $q_\mu=p_\mu-k_\mu$ is the momentum transferred to the lepton-neutrino pair. The $B_{(s)}$ and $D_{(s)}^*$ meson masses are given by $M_{B_{(s)}}$ and $M_{D_{(s)}^*}$.
  
In order to determine the form factors on the lattice, we need to calculate suitable ratios of 3-point and 2-point correlation functions 
\begin{align}
        R^{\Gamma,\mu}_{B_{(s)}\to D_{(s)}^*}(t,t_\text{snk})&=\dfrac{C_{B_{(s)}\to D_{(s)}^*}^{3pt, \Gamma,\mu}(t,t_\text{snk},k)}{\sqrt{ C_{D_{(s)}^*}^{2pt}(t,k)C_{B_{(s)}}^{2pt}(t_{snk}-t,p)}}\sqrt{\dfrac{4E_{D_{(s)}^*}M_{B_{(s)}}\sum_{j}\varepsilon_j(k)\varepsilon^{*j}(k)}{e^{-E_{D_{(s)}^*}}e^{-M_{B_{(s)}}(t_\text{snk}-t)}}}\nonumber\\
        &\qquad\xrightarrow[t_\text{snk}-t\rightarrow \infty]{t\rightarrow \infty}\varepsilon^\mu (k)\langle D_{(s)}^*(k, \varepsilon)|\overline{c}\Gamma b|B_{(s)}(p)\rangle\, ,
    \label{eq:ratio}
\end{align}
which in the limit of large Euclidean time separation between the $B_{(s)}$ and $D_{(s)}^*$, converge to the hadronic matrix elements of interest, and thus allow us to extract the lattice form factors from the following combinations of hadronic matrix elements
\begin{align}
    \widetilde{V}(q^2)&=-\frac{\text{i}}{2}\frac{M_{B_{(s)}}+M_{D_{(s)}^*}}{M_{B_{(s)}}}\frac{1}{k^n}\epsilon^{0ljn}\sum_\lambda\varepsilon_j(k,\lambda)\langle D_{(s)}^*(k,\lambda)|c\gamma_l b|B_{(s)}(p)\rangle\label{eq:ffv},\\
    \widetilde{A}_0(q^2)&=\frac{1}{2}\frac{M_{D_{(s)}^*}}{E_{D_{(s)}^*}M_{B_{(s)}}}\frac{1}{k_j}q^l\sum_\lambda\varepsilon_j(k,\lambda)\langle D_{(s)}^*(k,\lambda)|c\gamma_l\gamma_5 b|B_{(s)}(p)\rangle\label{eq:ffa0},\\
     \widetilde{A}_1(q^2)&=\frac{1}{M_{D_{(s)}^*}+M_{B_{(s)}}}\sum_\lambda\varepsilon^l(k,\lambda)\langle D_{(s)}^*(k,\lambda)|c\gamma_l\gamma_5 b|B_{(s)}(p)\rangle\label{eq:ffa1},\\
     \widetilde{A}_2(q^2)&=
     \frac{M_{D_{(s)}^*}^2\left(M_{B_{(s)}}+M_{D_{(s)}^*}\right)}
          {k_j^2 E_{D_{(s)}^*}M_{B_{(s)}}}
     \frac{q^2}{q^2+ M_{B_{(s)}}^2 -M_{D_{(s)}^*}^2}
     \left[\frac{-2k_j^2 E_{D_{(s)}^*}M_{B_{(s)}}}{q^2 M_{D_{(s)}^*}}
     \widetilde{A}_0(q^2) \right.\nonumber\\
     &+\nonumber\left(M_{B_{(s)}}+M_{D_{(s)}^*}\right)
     \left(1+\frac{k_j^2}{M_{D_{(s)}^*}^2}+\frac{E_{D_{(s)}^*} M_{B_{(s)}} k_j^2}
       {M_{D_{(s)}^*}^2 q^2}\right) \widetilde{A}_1(q^2)\\
     &-\sum_\lambda \varepsilon^j(k,\lambda)\langle D_{(s)}^*(k,\lambda)|c\gamma_j\gamma_5 b|B_{(s)}(p)\rangle\Bigg] \, .\label{eq:ffa2}
\end{align}
In Eqs.~\eqref{eq:ffv}--\eqref{eq:ffa2} we identify components of the 3-vectors $\vec k$ or $\vec q$ using the notation $k_i$ or $q_i$ i.e.~$1/k_i$ refers to the inverse of the $i$-component of $\vec k$. No summation over indices is implied and $\epsilon$ is the totally anti-symmetric Levi-Civita symbol. The tilde over the form factors reminds us that lattice quantities still need to be renormalized.  
We calculate the renormalization factors using mostly nonperturbative renormalization~\cite{Hashimoto:1999yp, El-Khadra:2001wco}, i.e.~the renormalization factor for a heavy-light current is written as
\begin{align}
    Z^{hl}_{J_\mu}=\rho^{hl}_{J_\mu}\sqrt{Z^{ll}Z^{hh}},
\end{align}
where the flavor diagonal parts $Z^{ii}$ are calculated nonperturbatively on the lattice and $\rho_{J_\mu}^{hl}$ is a residual correction close to unity calculated using 1-loop perturbation theory.
Moreover, we use the $\rho^{hl}_{J_\mu}$ factor to apply an overall multiplicative blinding factor which is unknown to the persons carrying out the analysis.

\section{Lattice details}
Our calculation takes advantage of existing data \cite{Flynn:2023nhi} generated using six RBC-UKQCD ensembles of gauge field configurations \cite{Allton:2008pn,Aoki:2010dy,Blum:2014tka,Boyle:2017jwu, Boyle:2018knm} featuring dynamical (2+1)-flavors of Shamir domain-wall fermions (DWF) \cite{Kaplan:1992bt,Shamir:1993zy,Furman:1994ky} with the Iwasaki gauge action \cite{Iwasaki:1983ck} at three different lattice spacings of $a^{-1}=$ 1.78, 2.38, and 2.78 GeV. We restrict ourselves to analyzing $B_s\to D_s^*\ell\nu$ decays where we choose a Shamir DWF with close to physical mass for the strange quarks, simulate charm quarks using Möbius domain-wall fermions \cite{Brower:2012vk} optimized for heavy quarks \cite{Cho:2015ffa,Boyle:2016imm}, and use the relativistic heavy quark (RHQ) action \cite{ElKhadra:1996mp,Christ:2006us,Lin:2006ur} for simulating physical mass bottom quarks on relatively coarse lattices where all parameters are tuned nonperturbatively \cite{Aoki:2012xaa,Flynn:2023nhi}. 
Details on the used ensembles are summarized in Tab.~\ref{tab:setup}. In our calculation we use point sources and sinks for the strange quark, but apply iterative Gaussian smearing to sources for bottom and charm quarks. In the case of the Möbius DWF for charm quarks, we either simulate two charm-like masses bracketing the physical value (medium and fine ensembles) or simulate three ``lighter-than-charm'' quarks to perform a small extrapolation (coarse ensemble). We perform the statistical data analysis using jackknife resampling and start by first averaging, if applicable, all sources on each gauge field configuration. 
\begin{table}[t]
    \centering
    \resizebox{\textwidth}{!}{%
    \begin{tabular}{l|c c c c c c c c }
    \hline\hline
    & L/a & T/a & $a^{-1}$ / GeV & $am_l^{sea}$ &$am_s^{sea}$ &$am_s^{val}$ & $M_\pi/$ MeV & srcs $\times$ $N_{conf}$\\\hline\hline
       C1 & 24 &64 &1.7848 &0.005 &0.040 & 0.03224(18) &340 &1 $\times$ 1636\\
         C2 & 24& 64& 1.7848& 0.010& 0.040& 0.03224(18) & 433& 1 $\times$ 1419\\
         M1& 32 &64 &2.3833 &0.004 &0.030 &0.02477(18) &302 &2 $\times$ 628\\
         M2 &32 &64 &2.3833 &0.006 &0.030 &0.02477(18) &362 &2 $\times$ 889\\
         M3 &32 &64 &2.3833 &0.008 &0.030 &0.02477(18) &411 &2 $\times$ 544\\
             F1S &48 &96 &2.785 &0.002144& 0.02144&0.02167(20) &268& 24 $\times$ 98  \\\hline\hline     
         \end{tabular}
         }
         \caption{RBC/UKQCD coarse (C), medium (M) and fine (F) gauge field ensembles with 2+1 flavour Shamir domain-wall fermions and Iwasaki gauge action \cite{Allton:2008pn,Aoki:2010dy,Blum:2014tka,Boyle:2017jwu, Boyle:2018knm}. } 
         \label{tab:setup}
\end{table}

\begin{figure}[tb]
    \includegraphics[width=0.45\linewidth]{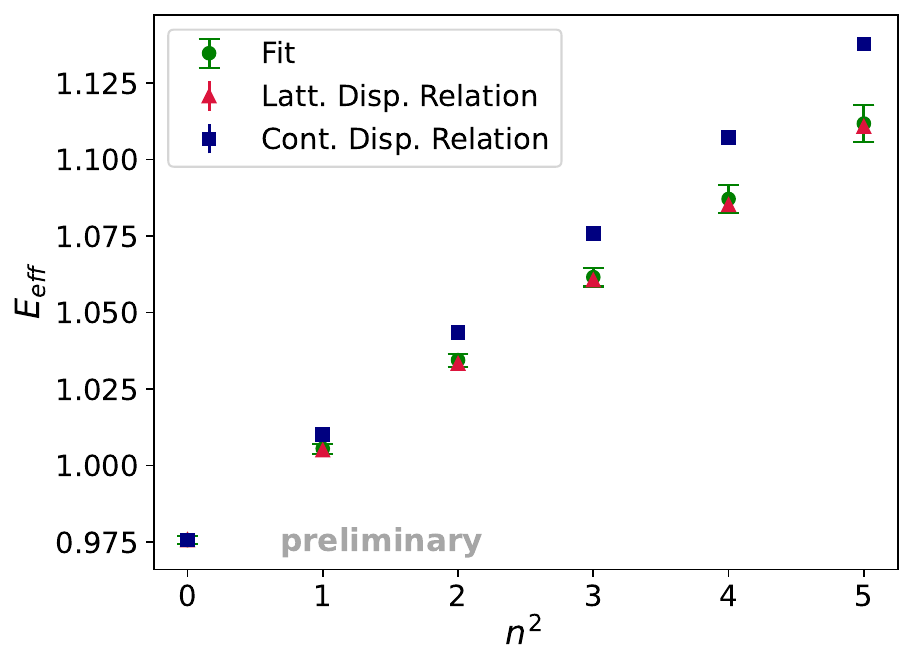}\hfill
    \includegraphics[width=0.45\linewidth]{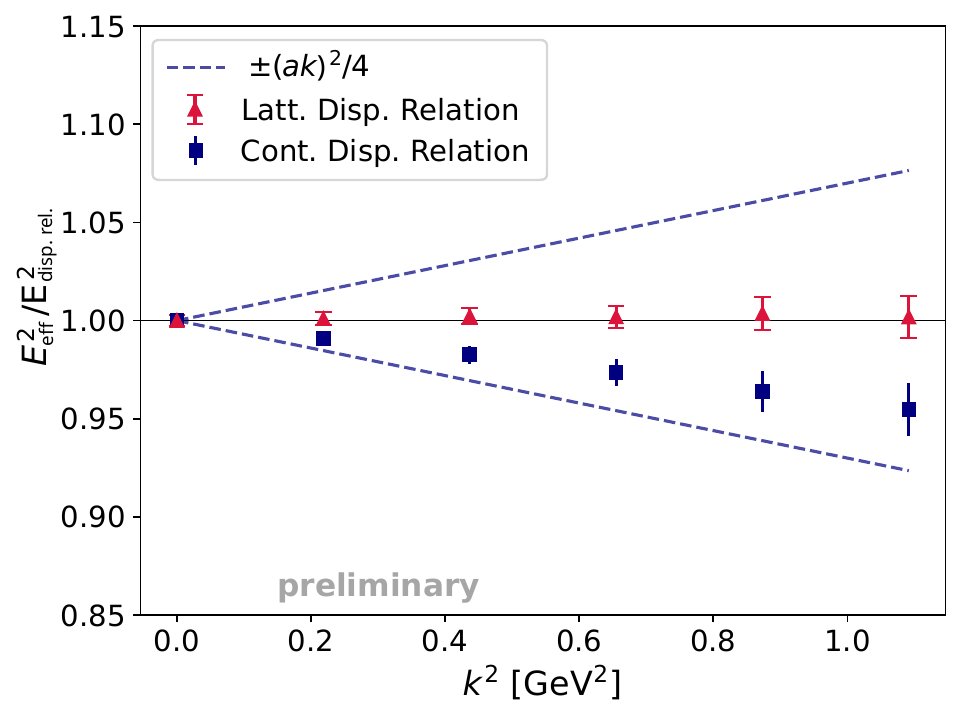}
    \caption{In the left panel we show the $D_s^*$ effective energies on C1, $am_c=0.300$ against the $D_s^*$ momentum in lattice units $n^2$. Green circles show the fitted values, red triangles are obtained using the lattice dispersion relation and blue squares using the continuum dispersion relation. In the right panel we plot the squared ratio of the fitted $D_s^*$ effective energies over the effective energies obtained using dispersion relations vs.~$k^2 = (2\pi n/L)^2$. The dashed lines indicate the leading order discretization error.}
    \label{fig:disprelc1}
\end{figure}

Before turning to the more complicated determination of form factors, we perform a first simple test of our setup using heavy Möbius DWF to simulate charm quarks. We calculate the energy of $D_s^*$ mesons injecting momentum ${\vec k}^{\,2}= (2\pi \vec{n}/L)^2$, with $\vec n^{\,2}=0,\,1,\,\ldots,\,5$, to check for discretization effects. Using our coarse ensemble C1 with $a^{-1}=1.7848\,\text{MeV}$, $L/a=24$, we show the outcome (green circles) in the left panel of Fig.~\ref{fig:disprelc1}.  In addition we use the rest mass (${\vec k}^{\,2}=0$) in combination with dispersion relations to check for consistency of the outcome. The blue squares are obtained using the continuum dispersion relation
\begin{align}
  E(\vec{k})=\sqrt{(am)^2+(a \vec{k})^2}\, ,
\end{align}  
while the red triangles use the lattice dispersion relation
\begin{align}
    E(\vec{k})=2a^{-1}\sinh^{-1}\sqrt{\sinh^2\left(\frac{am}{2}\right)+\sum_{i=0}^{3}\sin^2\left(\frac{ak_i}{2}\right)}\, .\label{eq:ldr}
\end{align}
Calculating squared ratios of our measured values over values obtained from a dispersion relation, we show the relative deviation in the right panel of Fig.~\ref{fig:disprelc1}. While the comparison to the continuum dispersion relation shows about 5\% deviation, which is below the naive power counting estimate indicated by the dashed lines, using the lattice dispersion results shows consistency within our statistical uncertainty.

\section{Extracting form factors}
\begin{figure}[tb]
    \includegraphics[height=0.24\textheight]{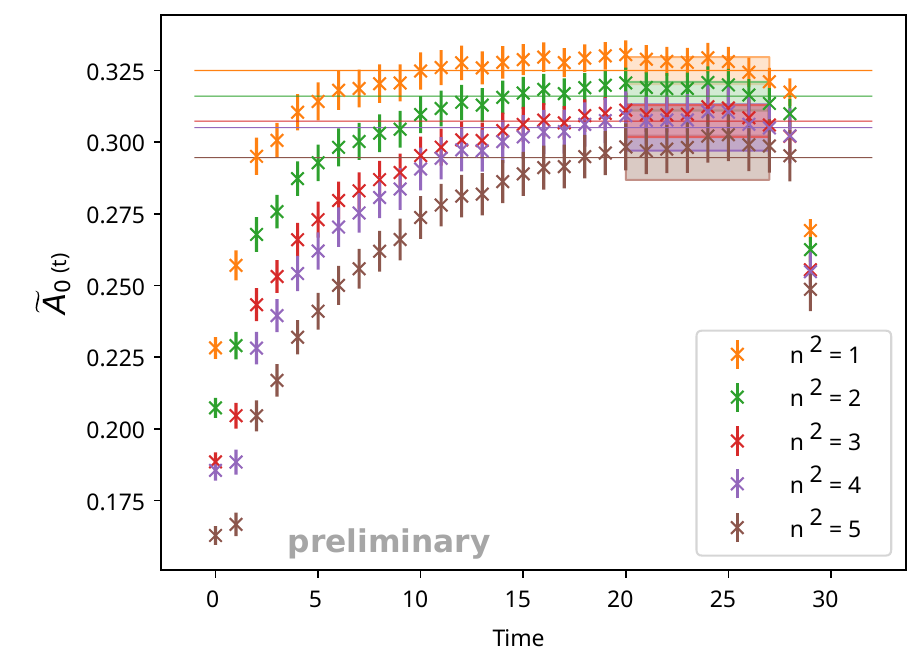}\hfill
    \includegraphics[height=0.24\textheight]{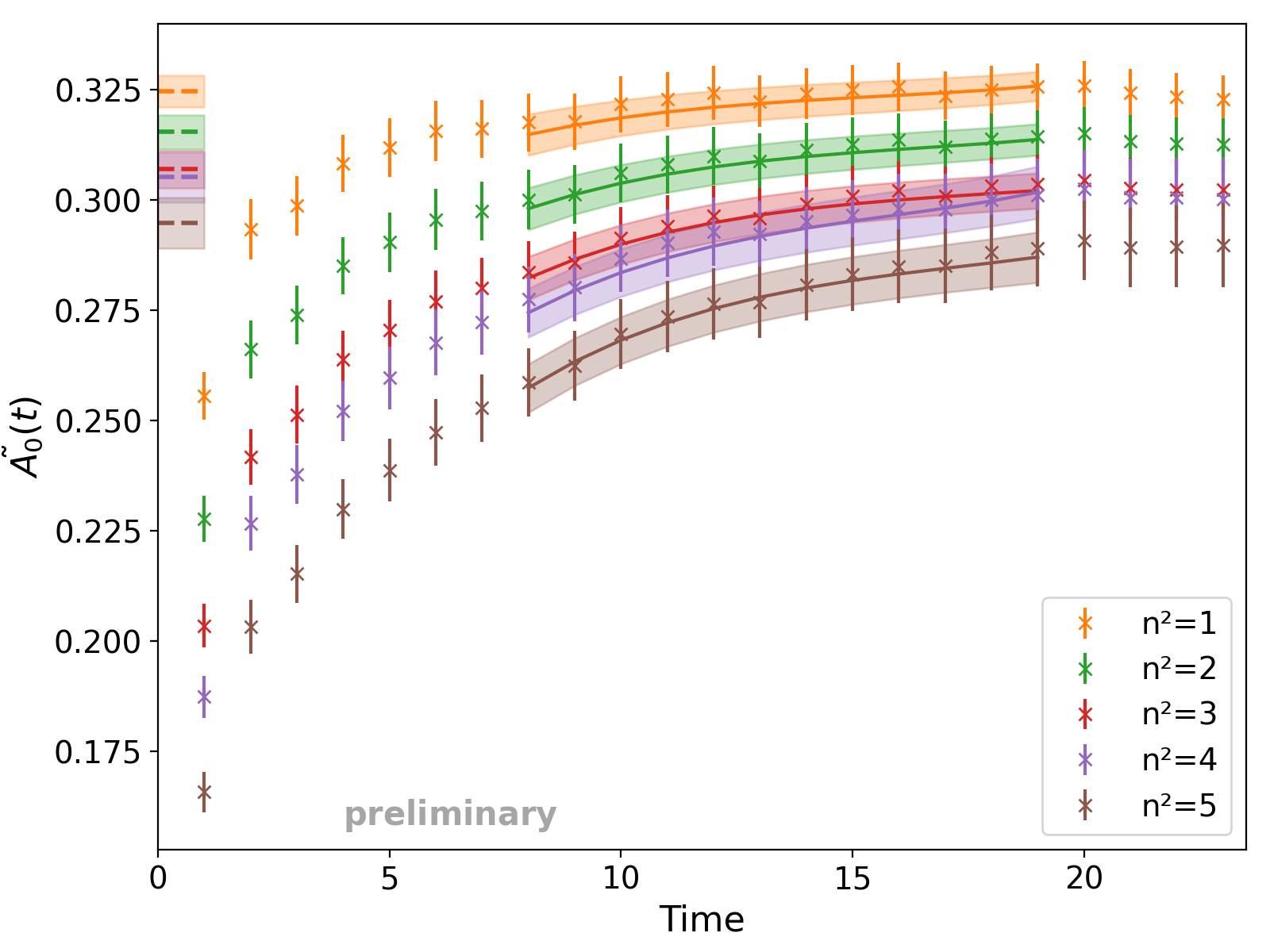}
    \caption{Comparison of extracting the signal for the $\widetilde A_0$ form factor on the F1S ensemble by performing a simple ground state fit (left) vs.~accounting for excited states (right). }
    \label{fig.compareA0}
\end{figure}

\begin{figure}[tb]
    \includegraphics[width=0.45\linewidth]{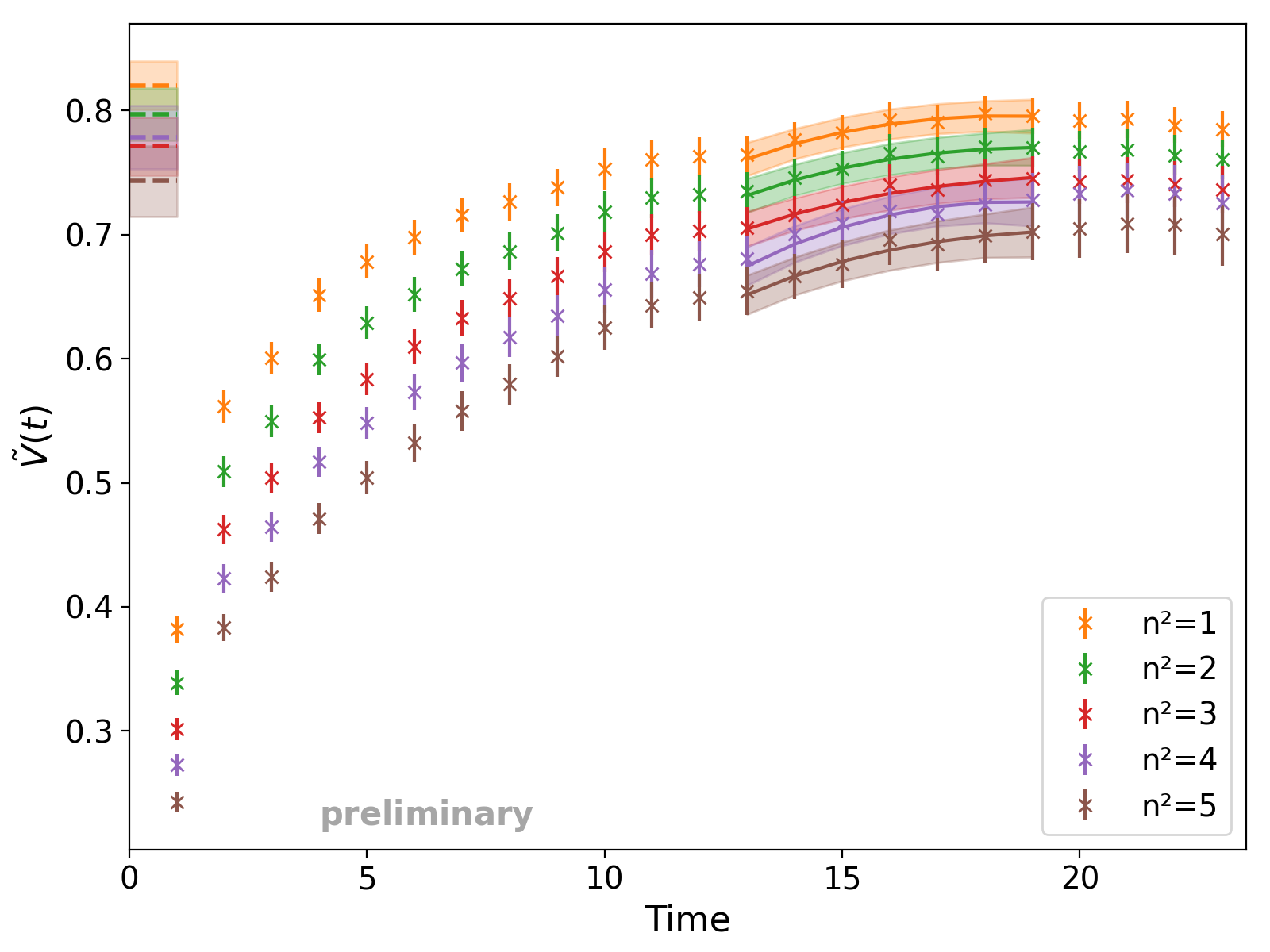}\hfill
    \includegraphics[width=0.45\linewidth]{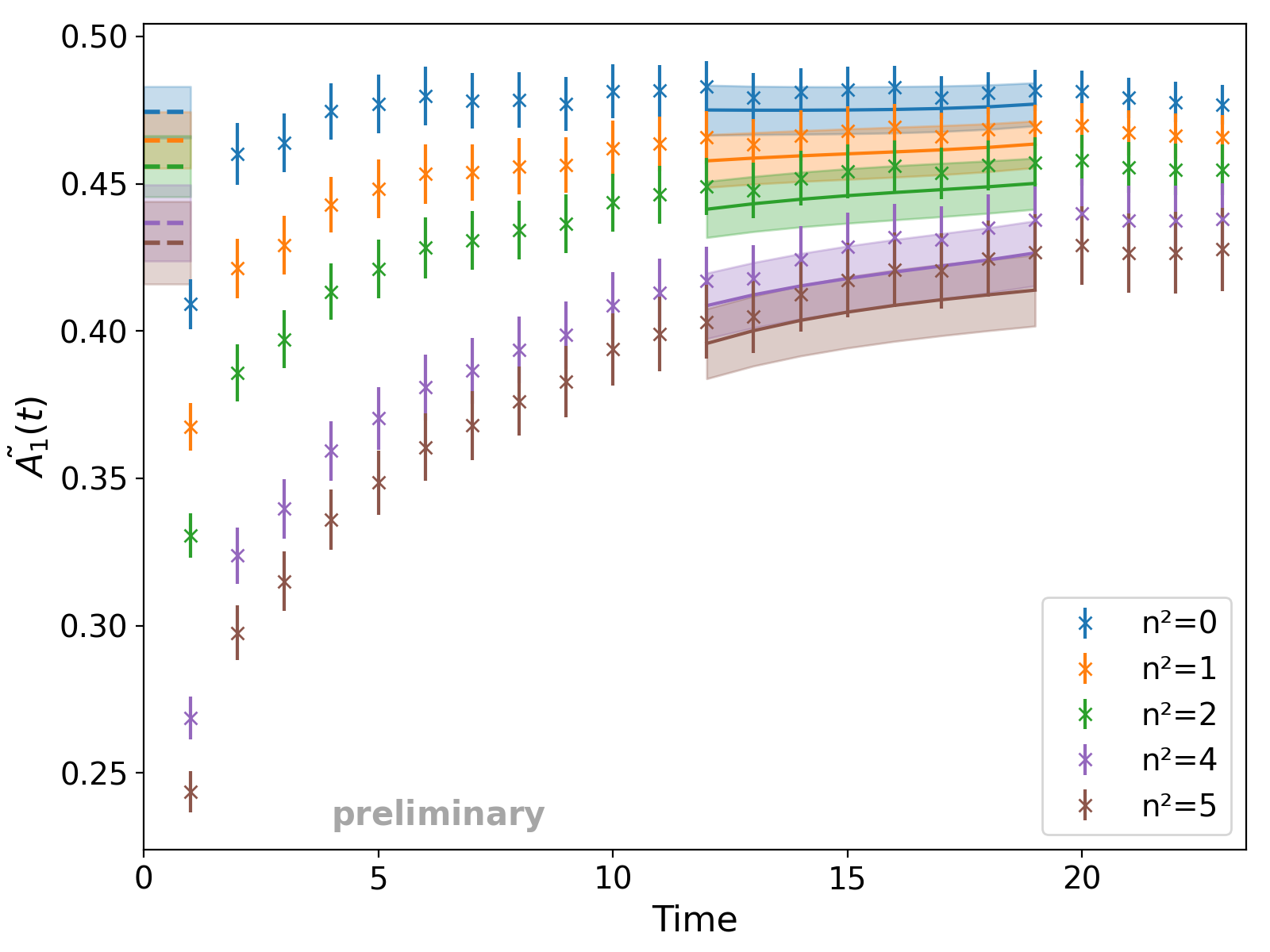}
    \caption{Extracting form factors $\widetilde{V}$ and $\widetilde{A}_1$ for F1S, $am_c=0.259$ projecting the $D_s^*$ hadronic final states to different momenta $\vec k^{\,2} = (2\pi \vec n/L)^2$ and performing a combined fit for all momenta parametrizing excited state contributions.}
    \label{fig:combinedexfits}
\end{figure}

The determination of the form factors starts by calculating ratios according to Eq.~\eqref{eq:ratio} for each ensemble and 
all measured final state momenta. For now we focus on $\widetilde A_0$, $\widetilde A_1$, and $\widetilde V$ because these three form factors allow for a direct extraction from specific matrix elements and are statistically more precise than $\widetilde A_2$. Moreover, our analysis is currently based on data obtained on four of the six ensembles listed in Tab.~\ref{tab:setup}. We highlight the importance of accounting for excited state contributions by showing for the F1S ensemble ($a^{-1}=2.785\,\text{GeV}$) the determination of $\widetilde A_0(t)$ in Fig.~\ref{fig.compareA0}. While the left panel exhibits some ambiguity which range of time slices to fit for extracting the form factor values, the inclusion of excited state contributions allows us to substantially enlarge the fit range, thus increasing the confidence in the extracted values. In both cases we perform one combined fit for all form factors at different $n^2$ and obtain good $p$-values $> 5\%$. Next we carry out fits including excited states for $\widetilde V$ and $\widetilde A_1$ shown in Fig.~\ref{fig:combinedexfits}.

\begin{figure}[tb]
    \includegraphics[width=0.32\linewidth]{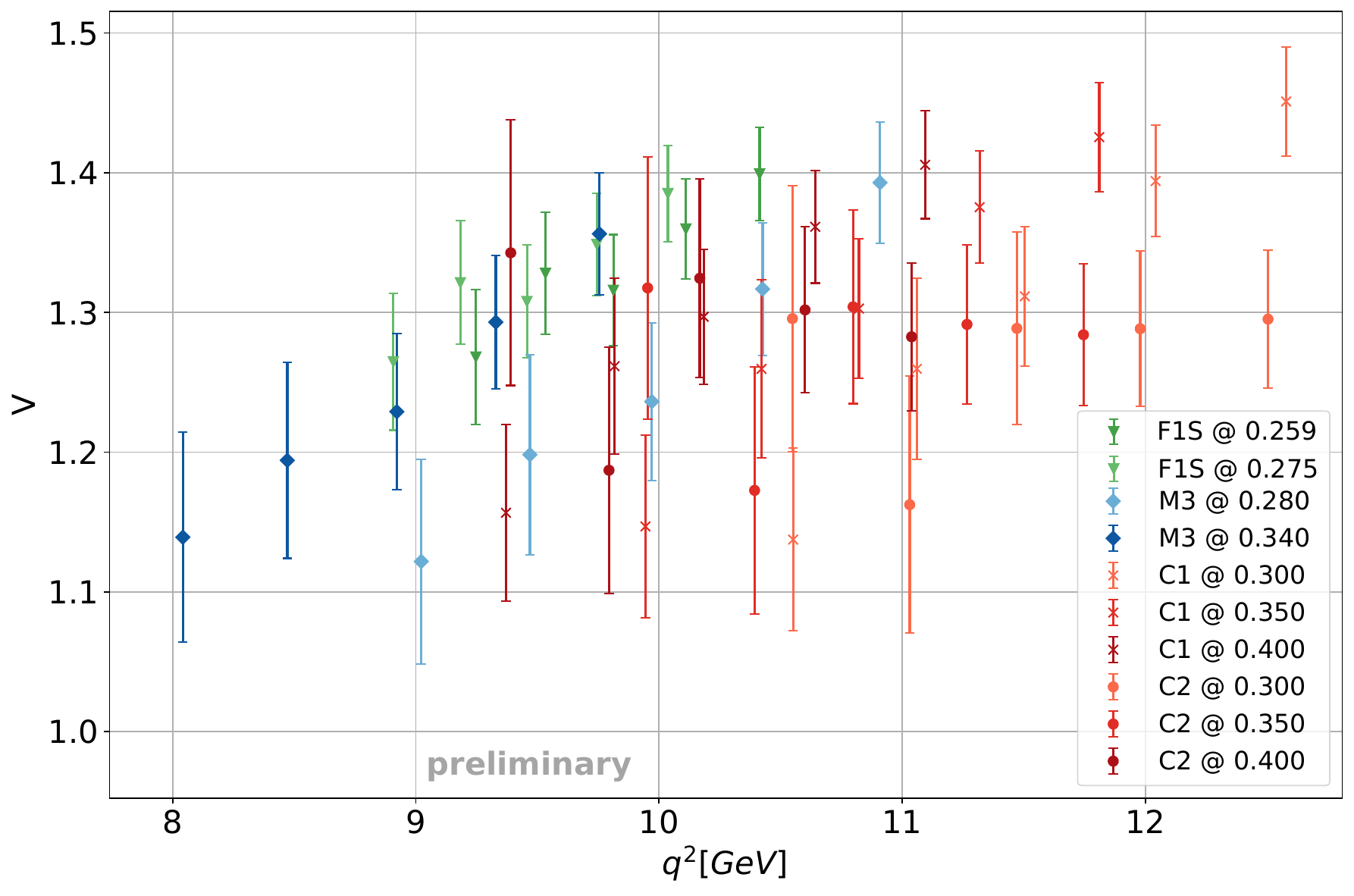}\hfill
    \includegraphics[width=0.32\linewidth]{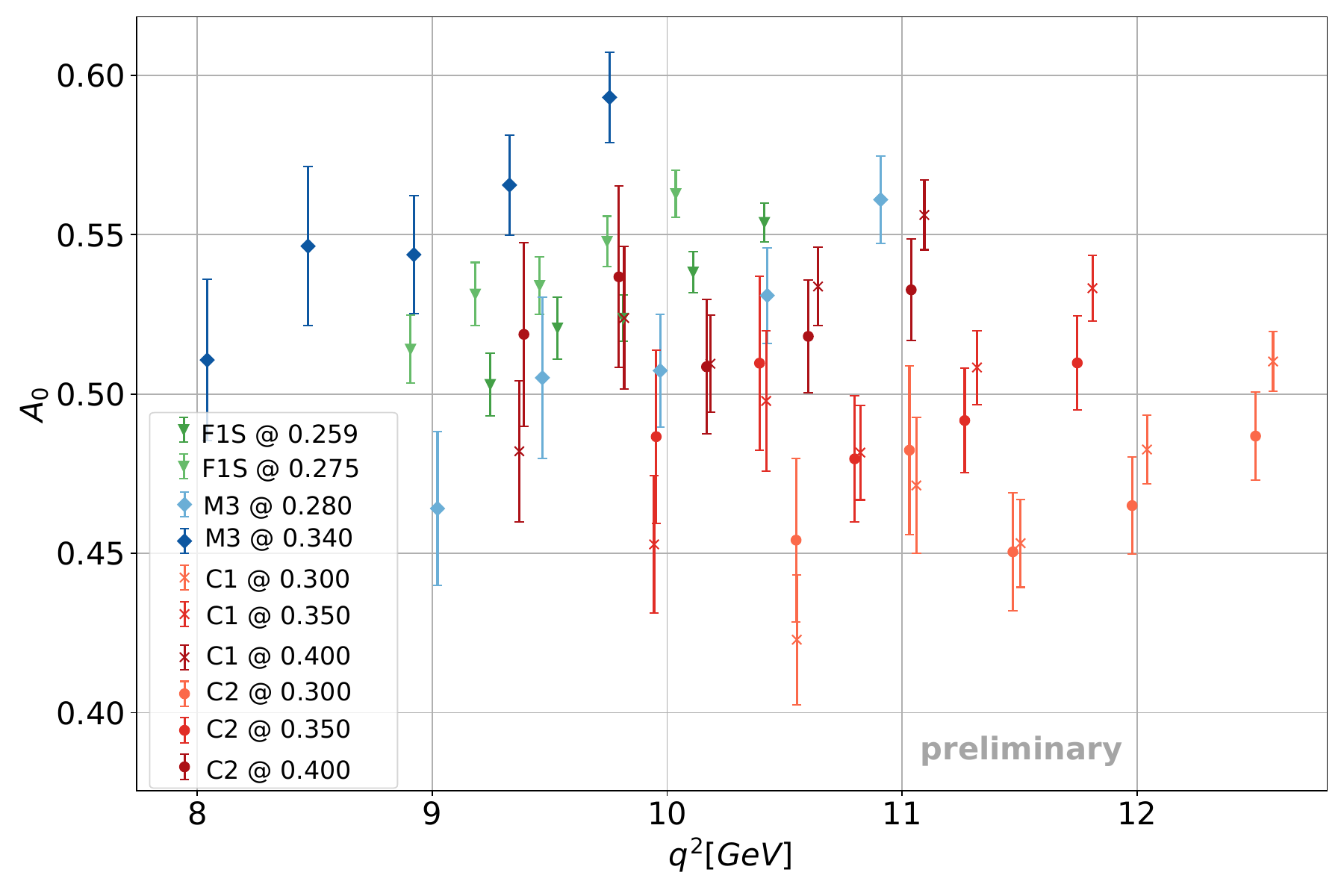}
    \hfill
    \includegraphics[width=0.32\linewidth]{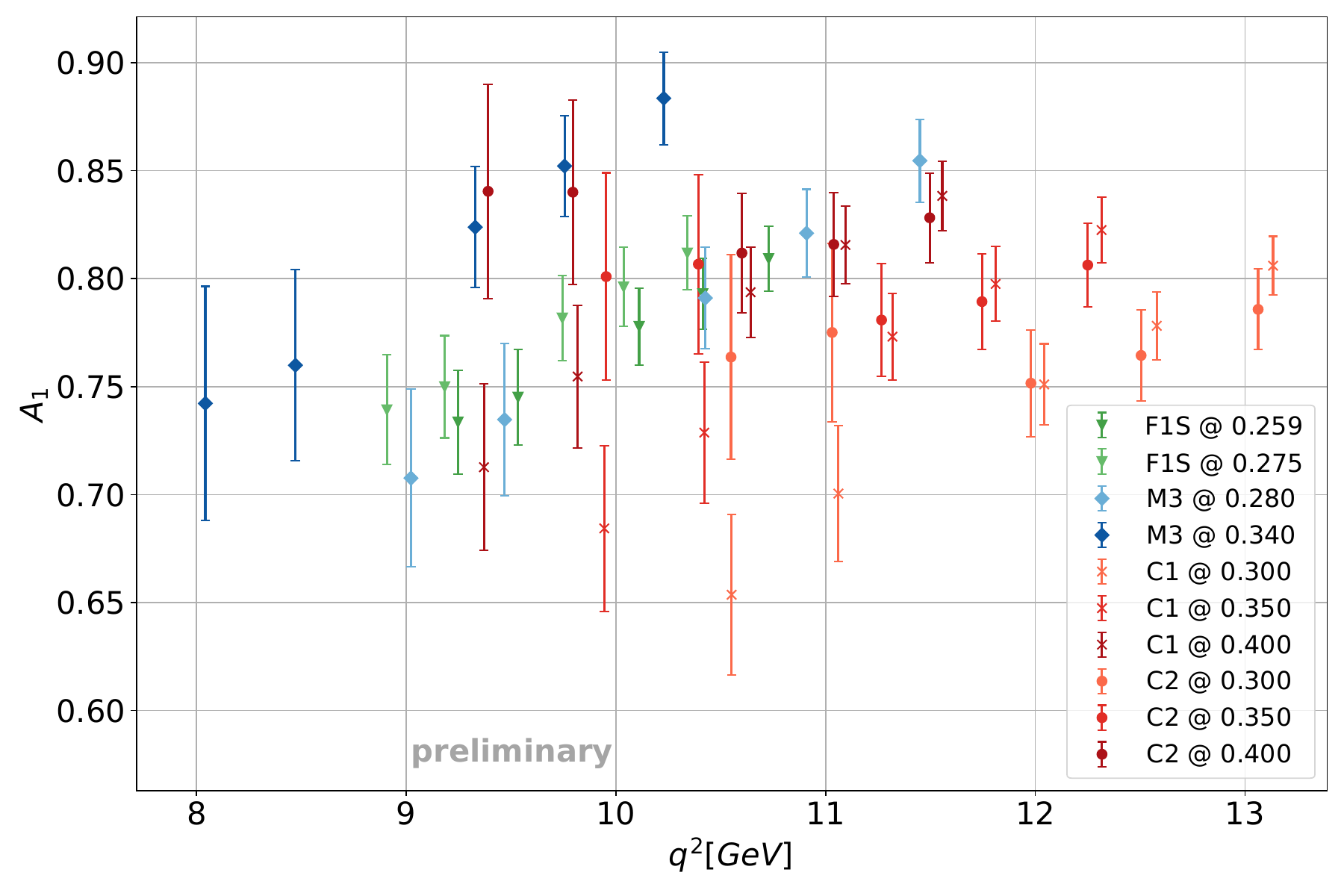}
    \caption{Blinded renormalized form factors $V, A_0$ and $A_1$ as a function of $q^2$. The different colors indicate the different ensembles and corresponding charm masses.}
    \label{fig:qsq}
\end{figure}

\begin{figure}[tb]
    \includegraphics[width=0.45\linewidth]{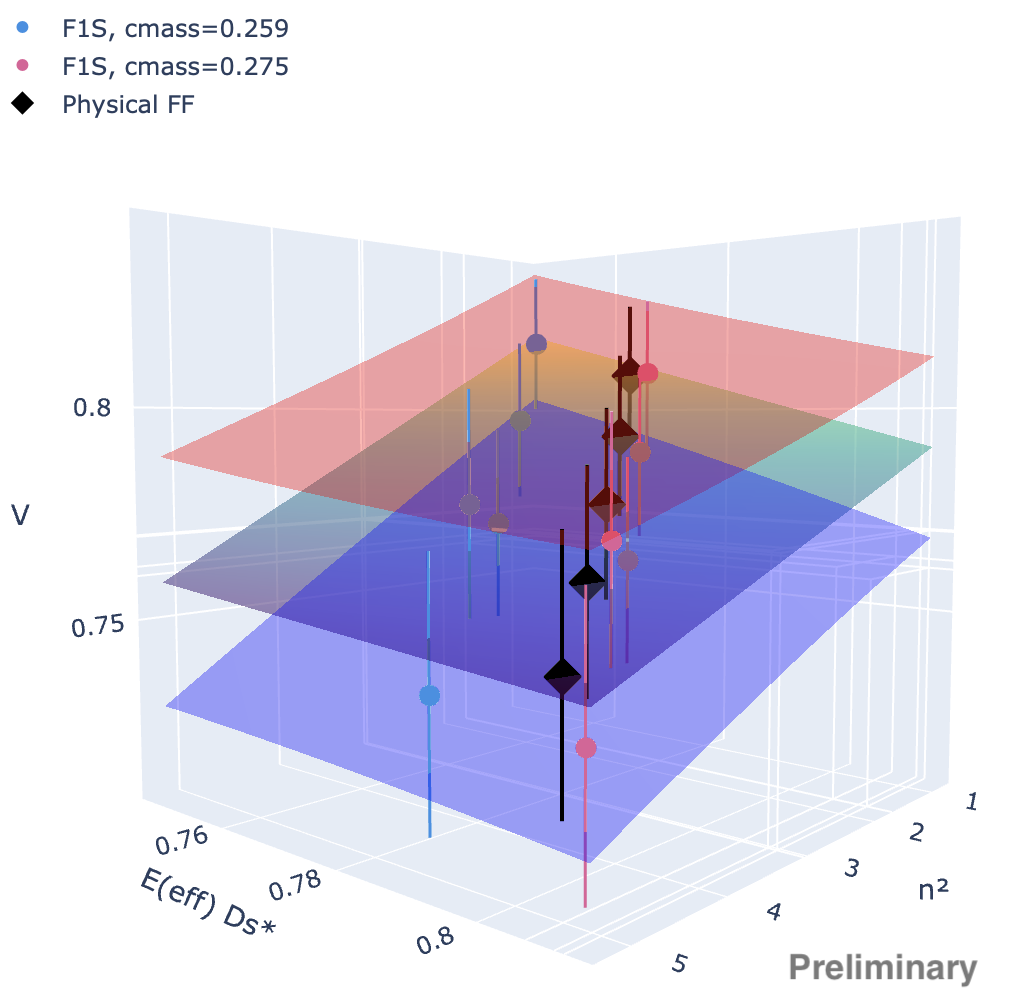}\hfill
    \includegraphics[width=0.45\linewidth]{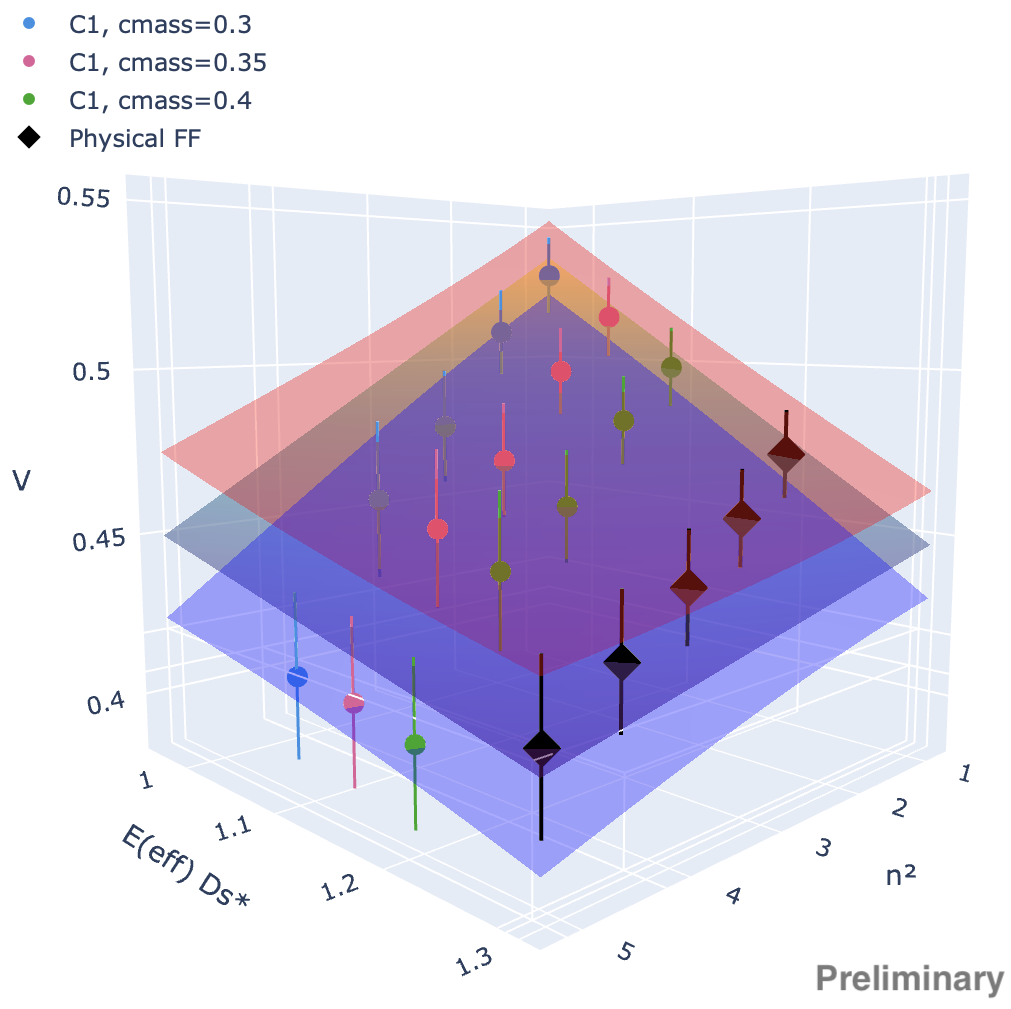}
    \caption{3-dimensional plots for the $V$ form factor on F1S (left) and C1 (right) with the $D_s^*$ momentum in lattice units $(2\pi \vec n/L)^2$ on the $x$-axis, the $D_s^*$ meson masses using the dispersion relation on the $y$-axis and the $V$ form factor fits on the $z$-axis. The central plane of the curved fit function defined in Eq. \eqref{eq:curved} in terms of $(2\pi \vec n/L)^2$ and $E_{\rm eff}^{D_s^*}$ is given by the green plane, while the $\pm 1\sigma$ planes are shown in red and blue. The values of the form factors corresponding to physical charm quark masses are shown by the black data points. }
    \label{fig:charm}
\end{figure}

After repeating this exercise for the other ensembles and charm-like quark masses, we multiply our blinded renormalization factors to plot the obtained form factors as a function of $q^2$ shown in Fig.~\ref{fig:qsq}. Next we need to combine our data in order to perform an extrapolation to the physical charm quark mass as well as perform a chiral-continuum limit. While in the end a ``global fit'' performing all extra- and interpolations at once may result in more precise results, we first prefer to keep more control over the different parts and perform a two step procedure. In the first step, we combine the data on one ensemble for  different momenta and charm-like quark masses, to  extrapolate (interpolate) to the physical charm quark mass on the coarse (medium and fine) ensembles. For this setup we perform a fully correlated fit using the ansatz
\begin{align}
    f(n^2,E_{\rm eff}^{D_s^*})=c_0+c_1 n^2+c_2E_{\rm eff}^{D_s^*}+c_3E_{\rm eff}^{D_s^*}n^2\, , \label{eq:curved}
\end{align}
which results in interpolations and extrapolations with $p$-values $>5\%$ on all ensembles. By construction the inter-/extrapolations depend on both the value of the charm quark mass as well as the units of momentum injected into the $D_s^*$ hadronic final state. Since we only fit data at the same value of the lattice spacing and spatial extent $L/a$ of the lattice, additional constants are for simplicity absorbed into the fit coefficients $c_i$. 
In Fig.~\ref{fig:charm} we show examples of extracting the $V$ form factor using an interpolation for the F1S ensemble on the left and an extrapolation for C1 on the right. In both cases the final interpolated/extrapolated values are shown by the black symbols and the colored planes intend to visualize the fitted plane and its $1\sigma$ uncertainties.

\begin{figure}[tb]
    \centering
    \includegraphics[width=0.33\linewidth]{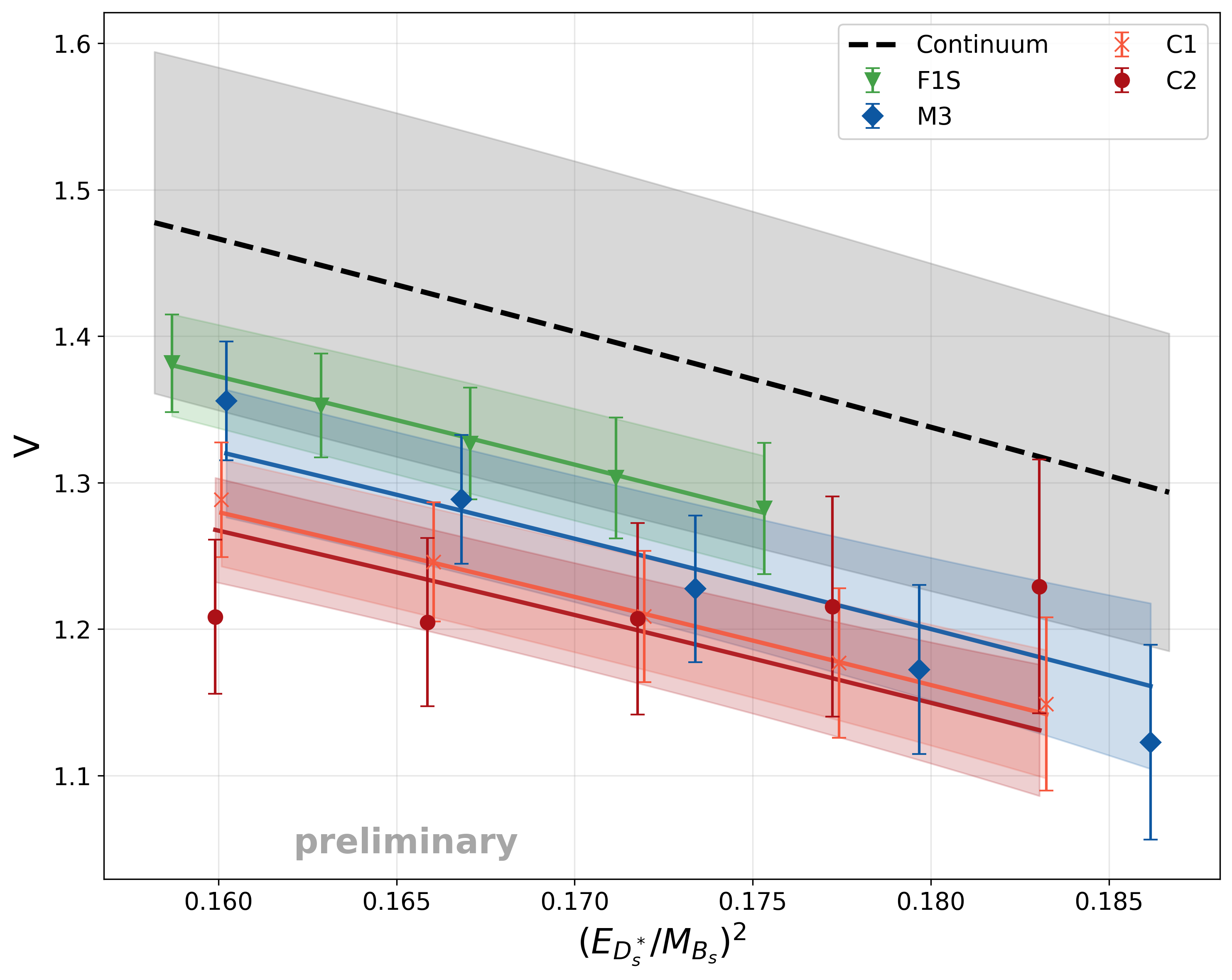}\hfill
    \includegraphics[width=0.33\linewidth]{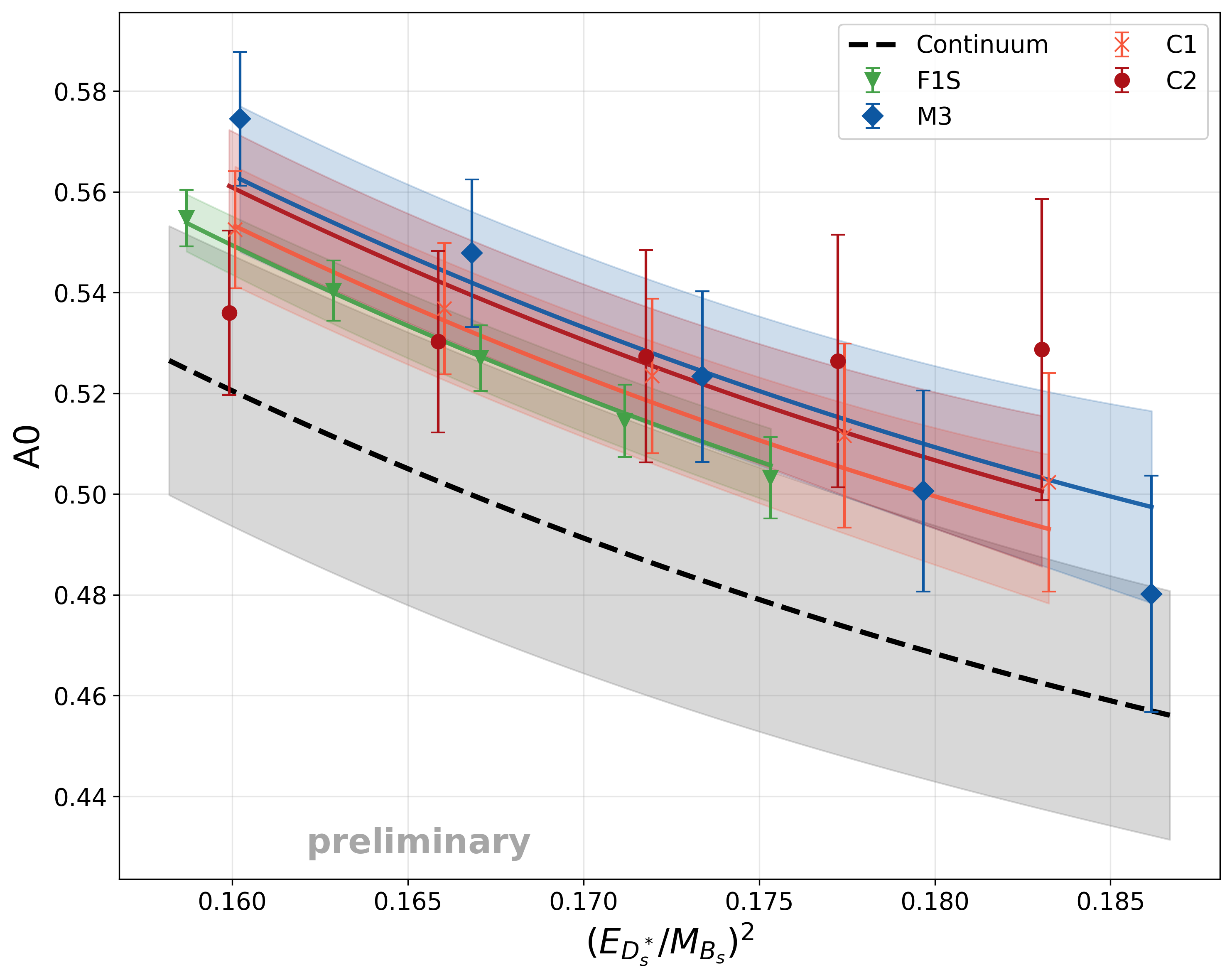}\hfill
    \includegraphics[width=0.33\linewidth]{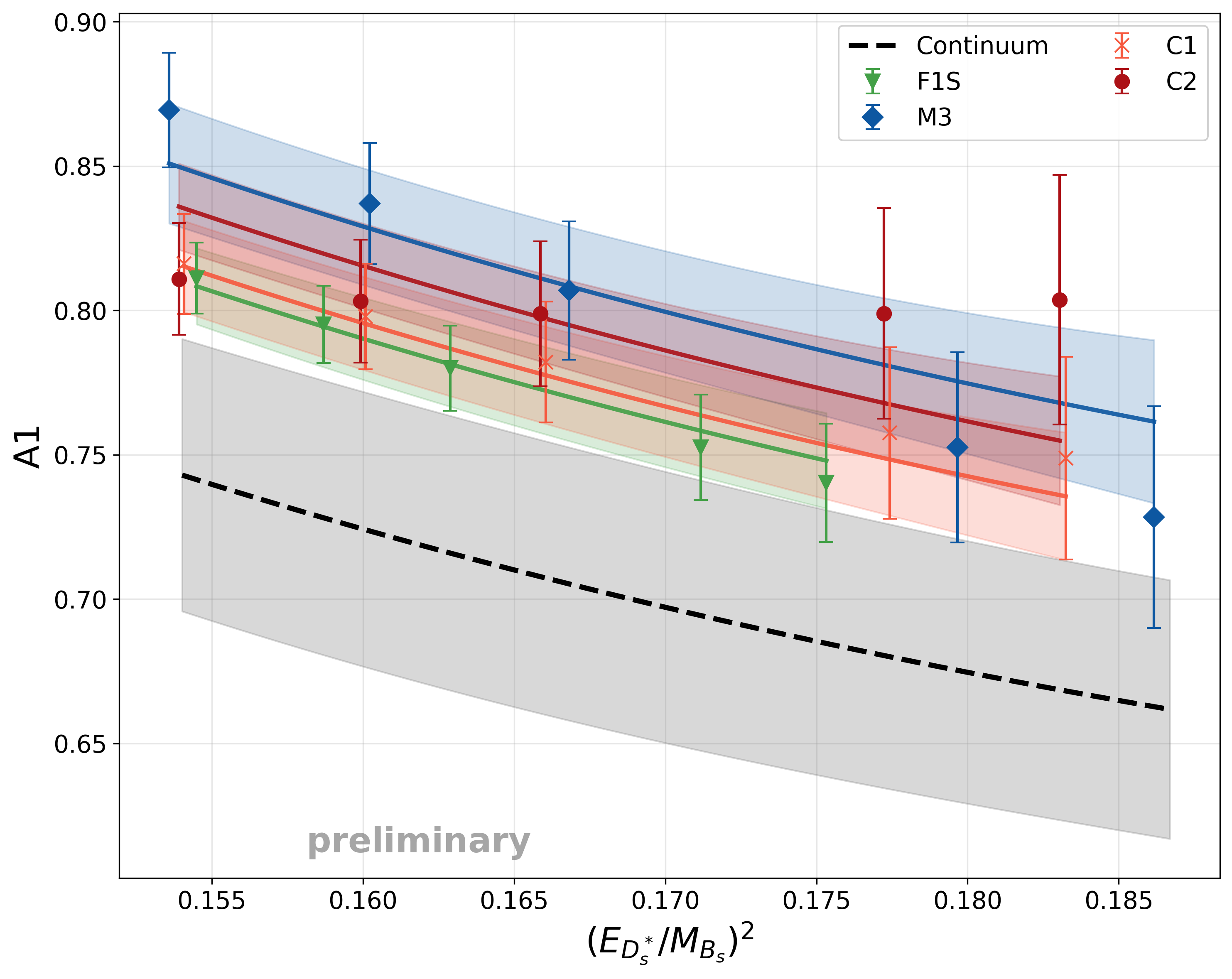}
    \caption{Chiral continuum fit defined in Eq. \eqref{eq:chiral} of the still blinded form factors $V, A_0$ and $A_1$ using four of intended six ensembles.}
    \label{fig:chiralwom}
\end{figure}

We use the resulting form factors as input for the second step performing a chiral-continuum extrapolation for which we choose the fit ansatz
\begin{align}
    f_{X}^{B_s\to D_s^*}\!\left(M_\pi,\,E_{D_s^*},\,a^2\right) &= \frac{\Lambda}{E_{D_s^*}+\Delta_X}\left[c_{X,0}+c_{X,1}\frac{\Delta M_\pi^{2}}{\Lambda^{2}} +c_{X,2}\frac{E_{D_s^*}}{\Lambda}+c_{X,3}\frac{E_{D_s^*}^{2}}{\Lambda^{2}} +c_{X,4}(a\Lambda)^2\right].
    \label{eq:chiral}
\end{align}
Our preliminary fit results are presented in Fig.~\ref{fig:chiralwom} where however only data obtained on four of the intended six ensembles have been included. 

\section{Summary}
Semileptonic $B_{(s)}\to D_{(s)}^*\ell\nu$ decays are phenomenologically interesting because these $b\to c$ transitions are related to two long-standing puzzles: the tension in $|V_{cb}^\text{incl}|$ vs.~$|V_{cb}^\text{excl}|$ as well as the tension in ${\cal R}(D^*)$. 
While $B^{0,+}$ meson decays are the most precisely measured by experiments, the determination of $B_s$ meson decays is favored on the lattice because the heavier strange quarks are numerically cheaper to calculate and more precise. 
Using existing RBC/UKQCD data, we investigate $B_s\to D_s^*\ell\nu$ semileptonic decays and extract the four form factors describing  decays to hadronic vector final states. 
Currently, an overall blinding factor is still applied and only data from four of six planned ensembles are included in our analysis. However, all steps of the analysis up-to the chiral continuum extrapolation are implemented and (blinded) results at physical quark masses are obtained. So far we have not accounted for any systematic uncertainties. 

\vspace{-10pt}
\acknowledgments 
This work was supported by the Deutsche Forschungsgemeinschaft (DFG, German Research Foundation) under Grant No.~396021762-TRR 257 ``Particle Physics Phenomenology after the Higgs Discovery''. AB ackowledges support from the House of Young Talents at the University of Siegen, Germany. M.B.\ was additionally funded in part by UK STFC grant ST/X000494/1.
Computations used resources provided by the USQCD Collaboration, funded by the Office of Science of the
U.S.~Department of Energy and by the \href{http://www.archer.ac.uk}{ARCHER} UK
National Supercomputing Service, as well as computers at Columbia University,
Brookhaven National Laboratory, and the OMNI cluster of the University of Siegen.
This document was prepared using the resources of the USQCD Collaboration at
the Fermi National Accelerator Laboratory (Fermilab), a U.S.~Department of
Energy (DOE), Office of Science, Office of High Energy Physics HEP User Facility. Fermilab is managed by Fermi Forward Discovery Group, LLC, acting under Contract No.~89243024CSC000002.
This work used the DiRAC Extreme Scaling service at the University of Edinburgh,
operated by the Edinburgh Parallel Computing Centre on behalf of the STFC
\href{https://dirac.ac.uk}{DiRAC} HPC Facility. This equipment was funded by BEIS capital
funding via STFC capital grant ST/R00238X/1 and STFC DiRAC Operations grant
ST/R001006/1. DiRAC is part of the National e-Infrastructure. 
We used gauge field configurations generated on
the DiRAC Blue Gene~Q system at the University of Edinburgh, part of the DiRAC
Facility, funded by BIS National E-infrastructure grant ST/K000411/1 and STFC
grants ST/H008845/1, ST/K005804/1 and ST/K005790/1.  

{
  \bibliography{B_meson.bib}
  \bibliographystyle{JHEP}
}
\end{document}